\documentclass[letter,twocolumn]{jpsj3}
\usepackage{color}
\usepackage{ulem}
\title{Partial Disorder and Metal-Insulator Transition in the Periodic Anderson Model on a Triangular Lattice} 

\author{Satoru HAYAMI\thanks{E-mail address: hayami@aion.t.u-tokyo.ac.jp}, Masafumi UDAGAWA, and Yukitoshi MOTOME 
}
\inst{Department of Applied Physics, University of Tokyo, 7-3-1 Hongo, Bunkyo-ku, Tokyo 113-8656, Japan 
}
\abst{
Ground state of the periodic Anderson model on a triangular lattice is systematically investigated by the mean-field approximation. 
We found that the model exhibits two different types of partially disordered states: 
one is at half filling and the other is at other commensurate fillings. 
In the latter case, the kinetic energy is lowered by forming an extensive network involving both magnetic and nonmagnetic sites, in sharp contrast to the former case in which the nonmagnetic sites are rather isolated. 
This spatially extended nature of nonmagnetic sites yields a metallic partially-disordered state by hole doping. 
We discuss the mechanism of the metal-insulator transition by the change of electronic structure. 
}

\kword{periodic Anderson model, geometrical frustration, triangular lattice, partial disorder, metal-insulator transition}

\begin{document}
\maketitle

Geometrical frustration is the conflict of competing interactions originating in the structure of underlying lattice~\cite{Diep200504,Lacroix2011introduction}.  
The competition leads to degeneracy in the low energy states, which is the source of fascinating phenomena, such as unusual orderings and exotic ground states. 
One of the peculiar orderings under geometrical frustration is a partial disorder (PD), which is characterized by coexistence of magnetic order and paramagnetic or nonmagnetic sites. 
This offers an intriguing example of self-organization to relieve geometrical frustration. 

PD was first theoretically proposed in the localized Ising spin model on a triangular lattice~\cite{Mekata1977antiferro-ferrimagnetic}, and later experimentally observed in $d$-electron compounds~\cite{Mekata1993successive,Niitaka2001partially}. 
In the localized spin systems, PD is mainly driven by the entropic effect, and hence it is stable only at finite temperatures. 
On the other hand, a metallic PD was also found in some $f$ and $d$-electron compounds~\cite{Mentink1994magnetic,Donni1996geometrically,Matsuda2012partially}. 
In stark contrast to the insulating PD, the metallic ones persist down to the lowest temperature, which suggests that itinerant electrons play a key role in stabilizing the PD states. 
In particular, in the $f$-electron compounds where itinerant electrons are coupled with localized $f$ moments, the Kondo singlet formation~\cite{Kondo1964resistance,Yosida1966bound} is expected to be relevant to the vanishment of moments at the nonmagnetic sites. 
In spite of extensive studies, the stabilization mechanism and the nature of these metallic PD states are not fully understood yet. 

In the present study, we theoretically examine the possibility of PD in itinerant electron systems, with focusing on the role of coupling between itinerant and localized electrons. 
Similar attempts were made for an effective pseudospin model~\cite{Dolores1997magnetic,Lacroix1996Kondo}, the Kondo necklace model~\cite{Motome2009instability,Motome2010partial,Motome2011variational}, the Kondo lattice model~\cite{Motome2010partial,Ishizuka2012}, and the periodic Anderson model~\cite{Hayami2011partial,Hayami2011carrier} on frustrated lattice structures. 
However, the obtained PD states have been all insulating so far, although metallic PD states are experimentally observed. 
It is highly desired to establish a theory to describe metallic PD states not only for understanding the experimental results but also for paving the way to explore unusual magnetotransport properties associated with the peculiar coexistence of magnetic and nonmagnetic sites. 

In this Letter, we systematically investigate the ground state of the periodic Anderson model on a triangular lattice. 
We identify new commensurate PD states, which are distinct from 
the one previously found at half filling~\cite{Hayami2011partial}. 
The striking feature is that the nonmagnetic sites contribute to the kinetic energy gain by participating in an extensive network with the magnetic sites, whereas they are rather isolated in the half-filling case. 
Reflecting the peculiar feature, we successfully obtain a metallic PD state by hole doping. 
We discuss the nature of the metal-insulator transition in detail. 

We consider the periodic Anderson model on a triangular lattice, whose Hamiltonian is given by
\begin{eqnarray}
\label{PAM_Ham}
{\mathcal{H}}  
\!\!\!\! &=&\!\!\!\! -  t \sum_{\langle i,j\rangle,\sigma} 
( c^{\dagger}_{i \sigma} c_{j \sigma}  + {\mathrm{H.c.}} ) 
- V\sum_{i ,\sigma} 
( c^{\dagger}_{i \sigma}f_{i \sigma}+{\mathrm{H.c.}} ) \nonumber  \\
 \!\!\!\!& &\!\!\!\!+U \sum_{i} f^{\dagger}_{i \uparrow}f_{i \uparrow} f^{\dagger}_{i \downarrow} f_{i \downarrow} 
+ E_{0} \sum_{i, \sigma} f^{\dagger} _{i \sigma} f_{i \sigma}. 
\end{eqnarray}
The first term represents the kinetic energy of conduction electrons, the second term the onsite $c$-$f$ hybridization between conduction $c$ and localized $f$ electrons, the third term the onsite Coulomb interaction for localized electrons, and the fourth term the atomic energy of localized electrons. 
The sum of $\langle i, j \rangle$ is taken over the nearest-neighbor sites on the triangular lattice. 
Hereafter, we take $t=1$ as an energy unit. 

In order to study the ground state of model Eq.~(\ref{PAM_Ham}), we employ the standard Hartree-Fock approximation for the Coulomb $U$ term, which preserves the SU(2) symmetry of the interaction term~\cite{Hayami2011partial,Hayami2011carrier}. 
In the calculations, we adopt a three-site unit cell to determine the global phase diagram. 
We also check the stability of the relevant phases by using a twelve-site unit cell, which accommodates longer-period orderings, such as period four ($2\times 2$) and six ($2\times 3$). 

We performed the calculations mainly at the commensurate fillings, $n=2m/3$ with an integer $m$ 
($n =\sum_{i \sigma} \langle c_{i \sigma}^{\dagger}c_{i \sigma} + f_{i \sigma}^{\dagger} f_{i \sigma}  \rangle/N$, where $N$ is the total number of sites), and the doped region around them. 
As a result, in addition to 
the PD state previously found at half filling ($n=2$)~\cite{Hayami2011partial}, we obtained new 
PD states at 2/3 and 8/3 fillings. 
In the following, we focus on the result at 2/3 filling as the nature of the PD state is similar to that at 8/3 filling. 

\begin{figure}[t!]
\begin{center}
\includegraphics[width=1.0 \hsize]{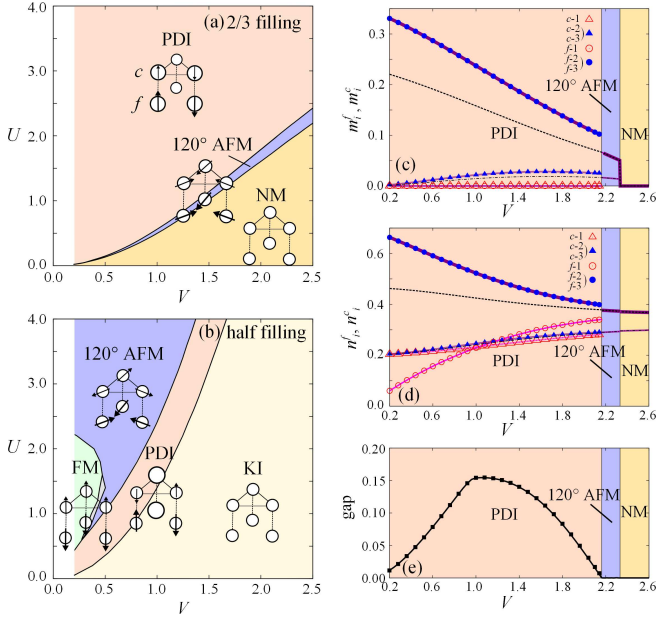}
\end{center}
\caption{
(color online). 
Ground-state phase diagram of the periodic Anderson model at (a) 2/3 filling and (b) half filling. 
$E_0$ is taken at $-4$ and $-1$, respectively. 
(b) is partly taken from ref.~\citen{Hayami2011partial} for comparison. 
Schematic picture of the ordering pattern is shown for each phase. 
The sizes of the circles reflect local electron densities, and the arrows represent local spin moments. 
PDI, AFM, NM, KI, and FM stand for the PD insulator, AF metal, nonmagnetic metal, Kondo insulator, and ferromagnetic metal, respectively. 
(c)-(e) $V$ dependences of (c) the magnitude of local spin moment, (d) local charge density, and (e) energy gap at 2/3 filling for $U=2$ and $E_0 =-4$. 
In (c) and (d), the data are plotted for $c$ and $f$ electrons at each sublattice separately, and the dot-dashed and dashed curves represent the mean values of $c$ and $f$ contributions, respectively. 
The solid lines (symbols) show the results of the calculations with the three-site (twelve-site) unit cell in (c)-(e). 
}
\label{Fig.1}
\end{figure} 

Figure~\ref{Fig.1}(a) shows the ground-state phase diagram at 2/3 filling. 
The phase diagram is divided into two regions: magnetically ordered region for $U \gtrsim V$ and nonmagnetic region for $U \lesssim V$. 
The magnetic and charge states as well as the energy gap are shown in Figs.~\ref{Fig.1}(c)-(e).  

In the region for $U \lesssim V$, the system is in a nonmagnetic metallic state. 
The magnetic moments vanish at all the sites in both $c$ and $f$ components [Fig.~\ref{Fig.1}(c)], and the system is a gapless metal [Fig.~\ref{Fig.1}(e)]. 
On the other hand, a 120$^{\circ}$ antiferromagnetic (AF) metallic phase appears in the narrow window next to the nonmagnetic metallic phase. In this phase, the $c$ and $f$ moments constitute a three-sublattice noncollinear 120$^{\circ}$ order, as schematically shown in Fig.~\ref{Fig.1}(a). 

While increasing $U$ or decreasing $V$, the system changes from metal to insulator with exhibiting a PD, as shown in Fig.~\ref{Fig.1}(a)~\cite{comment_large_U_region}. 
In this phase, one of the three sublattices becomes nonmagnetic, while the remaining two sublattices retain magnetic moments, which show a collinear AF order [Fig.~\ref{Fig.1}(c)]. 
The moments are larger for localized electrons than for conduction ones, and the magnitude of each moment takes the same value between the two sublattices. 
The PD is accompanied by charge disproportionation, as shown in Fig.~\ref{Fig.1}(d); 
the local charge density becomes higher at the magnetic sites than the nonmagnetic sites. 
The transition from the 120$^{\circ}$ AF metal to the PD insulator is of first order, while the charge gap grows continuously, as shown in Figs.~\ref{Fig.1}(c)-(e). 
We obtain essentially the same results by the calculations for the twelve-site unit cell, as plotted in Figs.~\ref{Fig.1}(c)-(e). 

\begin{figure}[t]
\begin{center}
\includegraphics[width=1.0 \hsize]{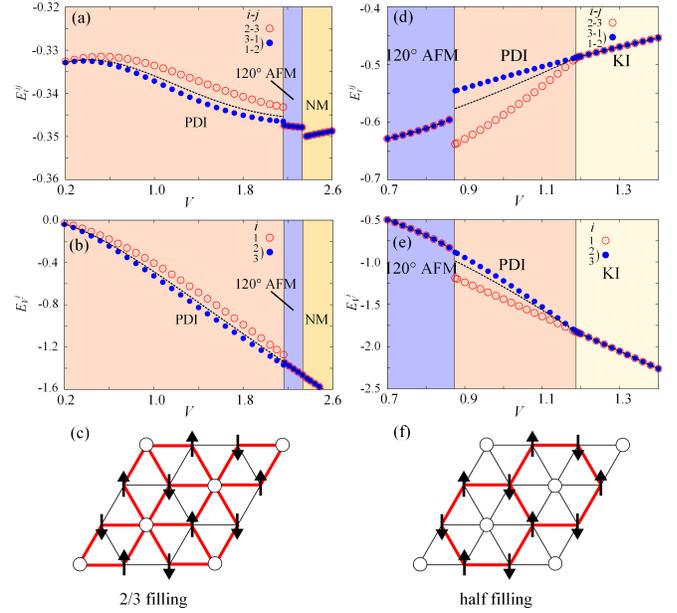}
\end{center}
\caption{
(color online). $V$ dependences of (a),(d) the kinetic energy for each bond and (b),(e) hybridization energy at each site. See the definitions in the text. 
The data are taken at (a),(b) $U=2$, $E_0=-4$, $n=2/3$ and (d),(e) $U=2$, $E_0=-1$, $n=2$.  
The symbols are common to those in Figs.~\ref{Fig.1}(c)-(e). 
(c),(f) Schematic pictures of the corresponding PD states; 
the thick lines show the bonds on which the 
kinetic energy gain is larger. 
}
\label{Fig.2}
\end{figure}

A similar three-sublattice PD insulating state was obtained by the authors at half filling between the 120$^{\circ}$ AF metallic phase and nonmagnetic Kondo insulating phase, as shown in Fig.~\ref{Fig.1}(b); the phase diagram is taken from ref.~\citen{Hayami2011partial} for comparison. 
However, there are several contrastive aspects between the two PD phases. 
First of all, the two PD states have different structures in the spin and charge ordering patterns.
As shown in the schematic pictures in Figs.~\ref{Fig.1}(a) and \ref{Fig.1}(b), the $c$ and $f$ moments at each magnetic site are aligned parallel to each other in the 2/3-filling case, while antiparallel in the half-filling case.
In addition, the local charge density is higher at the magnetic sites than the nonmagnetic site in the 2/3-filling case, whereas the tendency is opposite in the half-filling case. 

More interestingly, the two PD states show the distinct nature of spatial coherency. 
This is clearly observed by calculating the kinetic energy for nearest-neighbor bonds $\langle ij \rangle$, $E_{t}^{ij}=-t\langle \sum_{\sigma} c_{i \sigma}^{\dagger} c_{j \sigma} +\rm{H.c.} \rangle$, and the hybridization energy at site $i$, $E_{V}^{i}=-V \langle \sum_{\sigma} c_{i \sigma}^{\dagger}f_{i \sigma} +\rm{H.c} \rangle$. 
As shown in Fig.~\ref{Fig.2}(d), at half filling, the kinetic energy gain at the bond between a magnetic site and its neighboring nonmagnetic site is smaller than that between the magnetic sites in the PD state. 
This means that the nonmagnetic sites are weakly connected with the magnetic sites which form a honeycomb subnetwork of large kinetic energy gain, as schematically shown in Fig.~\ref{Fig.2}(f). 
Corresponding to this real-space separation, the $c$-$f$ hybridization gains a larger energy at the nonmagnetic sites than the magnetic sites, as shown in Fig.~\ref{Fig.2}(e).  
These indicate that the $c$-$f$ hybridization at the nonmagnetic site plays a crucial role in stabilizing PD at half filling.

In contrast, in the PD state at 2/3 filling, the nonmagnetic sites are not separated from the magnetic sites, but rather form a coherent network with magnetic sites, as schematically shown in Fig.~\ref{Fig.2}(c). 
This is clearly seen in the results plotted in Figs.~\ref{Fig.2}(a) and \ref{Fig.2}(b); the kinetic energy gain becomes larger for the bonds including nonmagnetic sites, and the hybridization energy gain becomes smaller for the nonmagnetic sites compared to the magnetic sites. 
In addition, the hybridization energy gain is overall smaller than that at half filling. 
These suggest less importance of the local $c$-$f$ hybridization for PD at 2/3 filling compared to half filling. 
This is reasonable since the average $f$-electron density at the nonmagnetic site is typically around 0.5, which is too small to gain the energy by forming a $c$-$f$ bonding state. 

The results above indicate that the nature of spatial coherency is clearly different 
between the PD states at half filling and 2/3 filling; 
the nonmagnetic sites are rather spatially isolated in the former, 
whereas they form an extended network with magnetic sites in the latter. 
This implies that, when going beyond the mean-field level, 
the nonmagnetic sites form spatially extended singlet states in the 2/3-filling PD state, 
whereas they form local Kondo singlets in the half-filling case.

\begin{figure}[t]
\begin{center}
\includegraphics[width=1.0 \hsize]{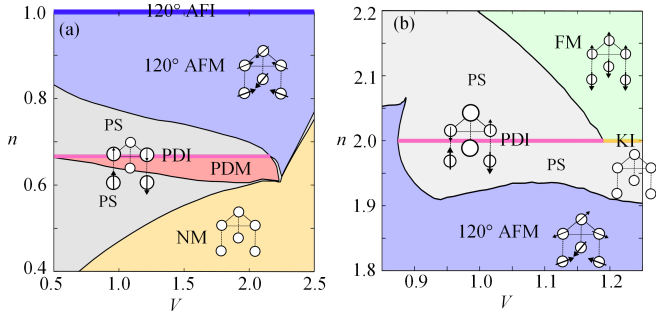}
\end{center}
\caption{
(color online). Ground-state phase diagram for carrier doping near (a) 2/3 filling and (b) half filling. 
The data are taken at (a) $U=2$, $E_0=-4$ and (b) $U=2$, $E_0=-1$. 
(b) is taken from ref.~\citen{Hayami2011carrier} for comparison. 
Schematic picture of the ordering pattern is shown for each phase. 
PDM and AFI stand for the PD metal and AF insulator. 
PS represents the phase separated region. 
In (a), the PDI and 120$^{\circ}$ AFI states exist only at $n = 2/3$ and $n=1$, respectively.
In (b), the PDI and KI states exist only at $n=2$. 
}
\label{Fig.3}
\end{figure} 

Reflecting the different coherency, the two PD insulators exhibit distinct responses to carrier doping. 
Figure~\ref{Fig.3} shows the results of the ground-state phase diagram for both hole and electron doping to 2/3 filling and half filling. 
The phase diagrams are obtained by using the grand canonical ensemble method~\cite{Hayami2011carrier}.  

In the 2/3-filling case, hole doping to the PD insulator induces a metal-insulator transition to a PD metallic state, as shown in Fig.~\ref{Fig.3}(a). 
The PD metallic state becomes stable in a wider range of doping while increasing $V$, whereas it disappears at $V\gtrsim 2.25$. 
On the other hand, the electron doping immediately leads to a phase separation to the 120$^{\circ}$ AF metallic state. 
The situation is quite different in the half-filling case; as shown in Fig.~\ref{Fig.3}(b) (taken from ref.~\citen{Hayami2011carrier} for comparison), both hole and electron dopings lead to a phase separation and no PD metallic state is obtained. 

\begin{figure}[t]
\begin{center}
\includegraphics[width=0.6 \hsize]{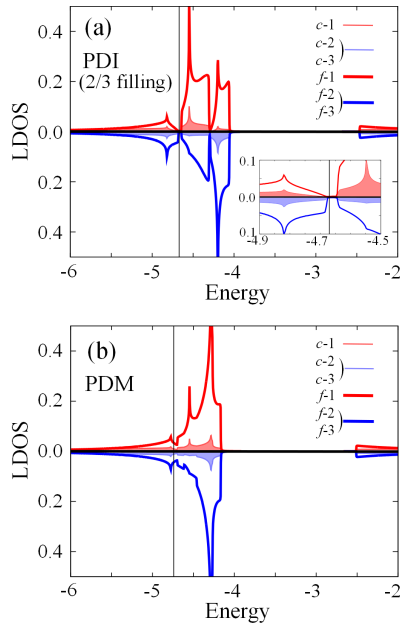}
\end{center}
\caption{
(color online). Local density of states of conduction and localized electrons for (a) PDI at $n = 2/3$ and (b) PDM at $n \simeq 0.617$. 
The data are taken at $U=2$, $V=2$, and $E_0=-4$. 
In the figures, the upper (lower) half shows the result for the nonmagnetic (magnetic) sublattice site(s) 1 (2 and 3). 
The data show the electronic structures near the Fermi level indicated by the thin vertical lines; the bands extend in the energy range approximately from $-7$ to $4$. 
In (a), the inset plots the enlarged figure in the vicinity of the Fermi level. 
}
\label{Fig.4}
\end{figure}

To understand the distinct responses to carrier doping between the half-filling and 2/3-filling cases, let us discuss the nature of the PD state and metal-insulator transition from the viewpoint of the electronic structure. 
Figure~\ref{Fig.4}(a) plots a typical local density of states for the PD insulating state at 2/3 filling. 
The results show the contributions from the nonmagnetic (upper half in the figures) and magnetic sites (lower half) separately. 
There is a small energy gap at the Fermi level, as shown in the inset of Fig.~\ref{Fig.4}(a). 
By hole doping, the gap closes as shown in Fig.~\ref{Fig.4}(b), while the PD magnetic structure persists even in the metallic state. 

The robustness of the PD metallic state near 2/3 filling can be attributed to the electronic structure near the Fermi level. 
In the 2/3-filling case, the density of states just below the Fermi level dominantly consists of the contribution from magnetic sites [inset of Fig.~\ref{Fig.4}(a)]. 
Hence, doped holes enter mainly into the magnetic sites. 
They reduce the ordered magnetic moments, but do not disturb the $c$-$f$ hybridized states at the nonmagnetic sites seriously; consequently, the PD state
persists as a metallic state upon hole doping. 
On the other hand, for electron doping, carriers are mainly doped into the nonmagnetic sites 
as the density of states just above the Fermi level is dominated by the nonmagnetic contribution [inset of Fig.~\ref{Fig.4}(a)]. 
In this case, the doped electrons directly disturb the $c$-$f$ hybridization, and hence, 
the PD state is unstable toward the phase separation to magnetically-ordered state.

In contrast, in the half-filling case, the density of states both just below and just above the gap is dominated by the contribution from nonmagnetic sites, i.e., the gap has a strong $c$-$f$ hybridization nature~\cite{Hayami2011partial}. 
In this case, both hole and electron dopings affect the nonmagnetic sites and reduce the energy gain by the $c$-$f$ hybridization. 
This gives rise to moment formation at the nonmagnetic sites, leading to the instability of PD state toward magnetically-ordered states. 

The above picture is consistent with the contrastive spatial coherency found in Fig.~\ref{Fig.2}. 
At half filling, doped carriers concentrate on the nonmagnetic sites which are rather separated from the magnetic subnetwork. 
It is not easy to make the system metallic by the doped $c$-$f$ hybridized states localized at the nonmagnetic sites; it seems necessary to reconstruct the magnetic order by inducing magnetic moments at the nonmagnetic sites, which immediately destroys PD. 
In contrast, the $c$-$f$ hybridized state is spatially extended including the magnetic sites at 2/3 filling. 
Carriers are not fatal to PD state, if they are mainly doped into the magnetic sites.
In this case, a metallic conductivity is easily acquired by the carriers doped on the extended coherent network. 
We expect that the PD metallic state remains stable even when we include quantum fluctuations beyond the mean-field approximation. 
This is because fluctuations may enhance the singlet formation by the c-f hybridization around the nonmagnetic sites and reduce the magnetic moments at magnetic sites, which is an opposite effect to a suppression of PD by inducing magnetic moments at the nonmagnetic sites. 
Such study beyond the mean-field approximation remains an interesting problem for future study.

To summarize, we have studied the ground state of the periodic Anderson model on a triangular lattice by the mean-field approximation up to the twelve-site unit cell. 
We have discovered partially disordered states at particular commensurate fillings, which are different from the previously-found half-filling one. 
In the new partially-disordered states, the nonmagnetic  states have a spatially extended nature and form an extensive network with magnetic sites. 
This is distinct from the situation at half filling, where the nonmagnetic sites are rather isolated from the magnetic honeycomb subnetwork. 
Reflecting the characteristic coherent nature, a metallic partially-disordered state is realized by carrier doping. 
The mechanism of the metal-insulator transition was analyzed from the viewpoint of the electronic structure. 
Our results will offer a good starting point to pursue the magnetotransport specific to the peculiar coexistence of magnetic and nonmagnetic sites.

\noindent {\bf Acknowledgements}\\
The authors acknowledge J. Yoshitake, Y. Akagi, and H. Ishizuka for fruitful discussions, and T. Misawa and Y. Yamaji for helpful comments. 
S.H. is supported by Grant-in-Aid for JSPS Fellows.
This work was supported by Grants-in-Aid for Scientific Research (No. 19052008, 21340090, 23102708, and 24340076), Global COE Program ``The Physical Sciences Frontier'', from the Ministry of Education, Culture, Sports, Science and Technology, Japan.  

\bibliographystyle{JPSJ}
\bibliography{ref}

\end{document}